\documentstyle[aps,prl,twocolumn]{revtex}
\begin{document}

\title{Proximity-induced time-reversal symmetry breaking at Josephson
junctions between unconventional superconductors}

\author{Kazuhiro Kuboki\cite{kuboki} and Manfred Sigrist}

\address{Department of Physics, Massachusetts Institute of Technology,
Cambridge, Massachusetts 02139}

\maketitle

\begin{abstract}

We argue that a locally time-reversal symmetry breaking state can
occur
at Josephson junctions between unconventional superconductors.
Order parameters induced by the proximity effect can combine with
the bulk order parameter to form such a state.
This property is specifically due to the intrinsic phase
structure of the pairing wave function
in unconventional superconductors.
Experimental consequences of this effect in high-temperature
superconductors are examined.
\end{abstract}


\narrowtext
A growing number of experiments demonstrates
that the superconducting state in some of
the high-temperature superconductors (HTSC) possesses
an unconventional order parameter which changes sign under
a $ 90^o $-rotation in the copper-oxide plane \cite{TEST}.
For a tetragonal system this state is commonly
refered to as a {\it d-wave}
state, where the Cooper pairing occurs in a channel
belonging to an irreducible representation of the
point group $ D_{4h} $, which is different from the
trivial representations, i.e. ``s-wave''.
Many of these experiments use the Josephson effect,
which allows a direct probe of the phase properties of a
superconducting order parameter.
While the Josephson effect in conventional (s-wave) superconductors
is described only by the phase difference between
the order parameters on both sides of an interface, the situation
is more complicated if the pair wave function has an intrinsic phase
structure like in the d-wave state. The sign change of the
d-wave state (generic pair wave function:
$ \psi ({\bf k} ) = \cos k_x - \cos k_y $)
between the $ x $- and $ y $-direction corresponds to a phase
difference of $ \pi $.
Due to this intrinsic phase structure,
the relative orientation of the pair wave function of two
connected superconductors and the geometry of the connecting interface
are important to determine
the phase relation between the order parameters on both
sides \cite{GESH,REV,YIP}. For d-wave superconductors
two types of Josephson junctions can occur, one where the interface
energy is minimized by a phase difference $ \varphi= 0 $
(standard 0-junction)
and the other with $ \varphi = \pi $ ($ \pi $-junction) \cite{WOHL}.
In multiply connected
superconducting systems the latter type can lead to frustration
effects and twists of the order parameter phase \cite{WOHL}. Resulting
phenomena
like the occurrence of spontaneous supercurrents and the modification
of standard interference patterns (SQUID) have
been exploited in experiments to
determine the symmetry of the order parameter
\cite{TEST}.

In this letter we add a new feature to the properties of
interfaces between unconventional superconductors.
We examine the effect of the reflection of Cooper pairs at the
interface and the mutual proximity of the two superconductors,
both of which can introduce Cooper pair amplitudes in
channels of symmetries different from that of the bulk
pairing state. These additional pairing components can
exist only in the immediate vicinity of the interface and
decay exponentially towards the bulk. However,
their presence has important consequences for the interface
properties because they modify the Josephson effect
via their influence on the current-phase relation.
The admixed order parameters originating from reflection and proximity
usually prefer a relative phase of 0 or $ \pi $ with the
bulk order parameters on both sides of the interface,
resulting in states conserving time-reversal symmetry.
We will show here that under certain conditions
the relative phase can change
leading to a state which breaks time-reversal symmetry at the
interface. It was shown by Sigrist, Bailey and Laughlin \cite{FRAC}
that this kind of interface state may
explain a recent experiment done by Kirtley et al.
where fractional vortices have been observed at boundaries
between differently oriented films of the HTSC
$ {\rm YBa}_2 {\rm Cu}_3 {\rm
O}_{7-\delta} $ (YBCO) \cite{KIRT}.

We discuss this problem using
the Ginzburg-Landau (GL) theory of a d-wave order parameter in a
tetragonal system ($ D_{4h} $) and
consider only one admixed order parameter (s-wave) which does
not occur in the bulk.
This choice is motivated by the experimental results
introduced above \cite{TEST} and the recent finding that the
critical behavior at the onset of superconductivity suggests
a single component bulk
order parameter in HTSC \cite{HARDY}.
The general GL free energy expansion for these
two components, $ \eta_d $ (d-wave) and $ \eta_s $ (s-wave),
in two dimensions
is given by

\begin{equation} \begin{array}{rl}
F = & \int d^2 x [\sum_{j=d,s} \{ a_j(T) |\eta_j|^2
+ \beta_d |\eta_j|^4 + K_j |{\bf D}
\eta_j|^2 \} \\
& \\
& \displaystyle + \gamma_1 |\eta_d|^2 |\eta_s|^2 + \frac{1}{2} \gamma_2
(\eta_d^{*2} \eta_s^2 + \eta_d^2 \eta_s^{*2}) \\
& \\
& + \tilde{K} \{ (D_x \eta_d)^* (D_x \eta_s) - (D_y \eta_d)^* (D_y
\eta_s) \\ & \\ &
\displaystyle + c.c. \} + \frac{1}{8 \pi} (\nabla \times {\bf A})^2 ]
\end{array}
\end{equation}

\noindent
with $ a_j(T) = \alpha_j (T - T_{cj}) $ and $ \alpha_j $,
$ \beta_j $, $ K_j $, $
\gamma_1 $, $ \gamma_2 $ and $ \tilde{K} $ are real coefficients
($ j=d,s $).
The gauge invariant gradient is given by $ {\bf D} = \nabla -i (2 \pi/
\Phi_0) {\bf A} $ with $ {\bf A} $ as the vector potential and
$ \Phi_0 = hc/2e $ is the standard flux quantum. We consider the
properties of an infinite planar
interface with the geometry shown
in Fig.1 where the crystal main ($ x $-) axis of the side A is pointing
perpendicular to the interface, while the orientation of the side B is
described by the angle $ \theta $ between the interface
normal vector and the crystal main ($ x $-) axis ($ -\pi/2
\leq \theta \leq \pi/2 $). In this arrangement
the effects of interest
occur only on side B. Thus, we will neglect the
detailed description of side A and represent it simply by a
complex (bulk) order parameter $ \eta_0 $. It can have d-wave
symmetry, but s-wave symmetry would not change any of our
conclusions qualitatively.
The properties of the interface are described by
additional interface terms in the GL free energy

\begin{equation} \begin{array}{ll}
F_{IF} =& \int_{IF} dS [ g_d (\theta) |\eta_d|^2 + \tilde{g}(\theta)
(\eta^*_d \eta_s + \eta_d \eta^*_s) \\
& \\
& + t_1(\theta) (\eta^*_0 \eta_d + \eta_0 \eta^*_d)
+ t_2 (\theta) (\eta^*_0 \eta_s + \eta_0 \eta^*_s) ]. \\
& \end{array} \end{equation}

\noindent
The first two terms describe the reflection properties
of the interface (IF) for the side B, i.e.: $ \eta_d \to \eta_d $
in the first and $ \eta_d \leftrightarrow \eta_s $ in the
second term (we neglect the $ \eta_s \to \eta_s $ contribution).
The latter two terms represent the lowest order
coupling between the two sides, A and B. By symmetry argument
we find that $ \tilde{g}(\theta \pm \pi/2) = - \tilde{g}(\theta) $
and $ t_1(\theta \pm \pi/2)= -t_1 (\theta) $. Thus, we choose
$ \tilde{g} = g_0 \cos (2 \theta) $ and $ t_1 =  t_0 \cos (2 \theta) $
with $ g_0, t_0 < 0 $ as a generic angle dependence.
We neglect any $ \theta $-dependence
in $ t_2 $ because none is required by symmetry ($ t_2 < 0 $).
The signs of $ g_0 $, $ t_0 $ and $ t_2 $ are chosen by
convention defining the specific gauge of the order parameter
phases. The coefficient $ g_d $ is in general positive
and describes the suppression of the d-wave order parameter at
the interface (see \cite{REV}).
The variational minimization scheme for the GL free energy
includes these interface terms into the boundary conditions.

This GL theory contains two competing tendencies for the
relative phase, $ \varphi_s - \varphi_d $, of the order
parameters $ \eta_j (= |\eta_j| {\rm exp}(i \varphi_j)) $,
($ j=s,d $). The last three terms in $ F_{IF} $ (Eq.(2)) are
minimized by $ \varphi_s - \varphi_d = 0 $ for $ |\theta| <
\pi/4 $ and $ \varphi_s - \varphi_d = \pi $ for $ \pi/4
< |\theta| < \pi/2 $. In either case, the state
at the interface is invariant under time-reversal
operation. On the other hand, if we assume
that $ \gamma_2 >0 $, the coupling
term $ \eta_1^{*2} \eta_2^2 + \eta_1^2 \eta_2^{*2} $ in Eq.(1)
favors energetically $ \varphi_s - \varphi_d = \pm \pi/2 $,
a state which breaks time-reversal symmetry (s+id-wave state).
(The assumption of positive $\gamma_2$ is natural,
since it would lead to a fully gapped state which is energetically
more favorable as weak-coupling studies of various microscopic
models show \cite{REV}.)
In the following we study the competition between the tendency
to keep or to break time-reversal symmetry near the interface
and some physical consequences.

Although our choice of geometry allows to reduce the problem
to one spatial dimension ($ \tilde{x} = x \cos \theta + y \sin \theta $
on side B), these GL equations can only be solved
numerically in general.
Fortunately, it is possible to understand the essential
properties of the competiting effects at the interface qualitatively
by using a simple variational ansatz for the order parameter.
We assume that as in HTSC the transition temperature $ T_{cd} $
of the d-wave order parameter is finite, while that
of the s-wave is very small or zero, $ T_{cs} \ll T_{cd} $
such that in the bulk only $ \eta_d $ is finite for $ 0 < T < T_{cd} $
(and $ \eta_0 $ on the side A) \cite{TEST,HARDY}.
In our variational treatment $ \eta_d $ shall
be real and independent of position on the side B,
and $ \eta_0 $ of the side A shall have a fixed modulus $ \tilde{\eta}
$ and a phase $ \chi $,
$ \eta_0 = \tilde{\eta} {\rm exp} ( i \chi) $
($ \eta_d(T), \tilde{\eta} (T) $ are identical to their bulk values).
Hence, this ansatz neglects the direct effect of the interface
on the modulus of the d-wave order parameter.
The s-wave component $ \eta_s $ is complex and decays
exponentially towards the bulk on the side B, $ \eta_s (\tilde{x}) =
\hat{\eta}
{\rm e}^{-\tilde{x}/\xi} $ and $ \hat{\eta} = \hat{\eta}'
+ i \hat{\eta}'' $.
The part of the free energy (per unit interface area)
depending on the variational degrees of
freedom, $ \xi $, $ \hat{\eta} $ and $ \chi $, is
obtained straightforwardly

\begin{equation} \begin{array}{ll}
F_{var} (\hat{eta} ,\xi, \chi) = & \displaystyle K_+
\frac{\hat{\eta}'^2}{2 \xi} + K_- \frac{\hat{\eta}''^2}{2 \xi} \\
& \\
& + 2 \{ \tilde{g} (\theta) \eta_d \hat{\eta}' + t_1 (\theta)
\tilde{\eta} \eta_d  \cos \chi \\
& \\
& + t_2 \tilde{\eta} (\hat{\eta}' \cos \chi
+ \hat{\eta}'' \sin \chi) \} \\
& \end{array}
\end{equation}

\noindent
with $ K_{\pm} = K_s + a_{\pm} \xi^2 $ and
$ a_{\pm} = a_s + (\gamma_1 \pm \gamma_2) \eta_d^2 $.
To be consistent with our assumption that $ \eta_s =0 $ in the
bulk we require $ a_{\pm} > 0 $. Further, we choose
$ a_- < a_+ $ ($ \gamma_2 > 0 $). By minimizing $ F_{var} $
we find that
$ \xi $ is always finite with
$ K_s/a_+ \leq \xi^2 \leq K_s/a_- $.
Some simple algebra leads to the following equation for
$ \chi $,

\begin{equation}
\sin \chi [\cos \chi + \frac{(t_1 K_+ - 2 \xi \tilde{g} t_2)
K_-}{4 t_2^2 \xi^3 \gamma_2 \tilde{\eta} \eta_d}] = 0.
\end{equation}

\noindent
One set of solutions of this equation is $ \chi = 0 $ and $ \pi $
($ \sin \chi = 0 $)
which leads to $ \hat{\eta}' \neq 0 $ and $ \hat{\eta}'' = 0 $
such that the relative phase between
$ \eta_d $ and $ \eta_s $ is 0 or $ \pi $. Energetically
$ \chi = 0 $ ($ \hat{\eta}'>0 $) is favored for $ |\theta|
< \pi/4 $ (0-junction)
and $ \chi= \pi $ ($ \hat{\eta}' <0 $)
for $ |\theta| > \pi/4 $ ($ \pi $-junction).
However, also the term in $ [...] $ can vanish giving an alternative
solution of Eq.(4) with $ \chi $ different from the above two limiting
values. Both $ \hat{\eta}' $ and $ \hat{\eta}'' $ are finite in
this case. This state is
two-fold degenerate because the application of the time-reversal
operation ($ \eta \to \eta^* $) corresponding to $ \chi \to - \chi $
and $ \hat{\eta} \to \hat{\eta}^* $ leads to a different
state with the same free energy.
Consequently, this solution violates time-reversal
symmetry $ {\cal T} $. It is easy to
see that the energy is indeed minimized by this state if
$ \gamma_2 $ is positive.
{}From Eq.(4) it is obvious that the $ {\cal T} $-violating
state can only occur if the modulus of the second term in $ [...] $
is smaller than 1. This condition can be satisfied by
an appropriate choice of the angle $ \theta $, because this
term is proportional to $ \cos(2 \theta) $ via
$ t_1 $ and $ \tilde{g} $, which can be arbitrarily small
for $ \theta $ close to $  \pm \pi/4 $.

{}From the above consideration the condition for the instability of
the time-reversal invariant state is obtained as

\begin{equation}
\cos (2 \theta)
= \pm \left| \frac{t_2^2 \gamma_2 \tilde{\eta} (T^*) \eta_d (T^*)}
{(a_s + \gamma_1 \eta_d^2)
(t_0 \sqrt{K_s a_+} - t_2 g_0)} \right|,
\end{equation}

\noindent
where $ T^* $ ($ < T_{cd} $)
denotes the critical temperature of the second order
transition to the $ {\cal T} $-violating state for a given
angle $ \theta $. The denominator on the right
hand side of this equation
is only weakly dependent on $ T^* $ close to
$ T_{cd} $ so that $ T^* $ is large for $ \theta $ near $ \pm \pi/4 $,
and is $ T_{cd} $ at
$ \theta = \pm \pi/4 $. We show a schematic phase diagram,
$ T $ versus $ \theta $, corresponding to
Eq.(5) (Fig.2). The $ {\cal T} $-violating phase
appears in a certain range of $ \theta $ around $ \pi/4 $ and
the region
of this phase becomes wider with decreasing temperature.
The specific shape of the phase boundary line depends on properties
of the system, in particular, of the interface.

Let us now investigate important consequences of the existence of
a $ {\cal T} $-violating interface state, which result from
the properties of its Josephson current-phase relation,

\begin{equation}
I=I_c g(\varphi)=I_c g(\varphi+2 \pi)
\end{equation}

\noindent
where $ \varphi $ is the phase difference between the order parameters
$ \eta_0 $ and $ \eta_d $ and $ g(\varphi) $ is a $ 2 \pi $-periodic
function of $ \varphi $ ($ |g(\varphi)| \leq 1 $) \cite{YIP}.
The specific form of $ g $ is not important here
and will be discussed elsewhere. The interface state with minimal
energy
corresponds to $ I=0 $ and its twofold degeneracy
implies that two functions, $ g_+ $ and $ g_- $, exist
with $ g_+(\chi)=0 $ and $ g_-(-\chi)=0 $, each belonging to one
of the two states. There are two possible
situations: I) $ g_+(\varphi) \neq g_-(\varphi) $ describing both
different
(meta)stable states as a function of $ \varphi $, and II)
$ g_+(\varphi) = g_-(\varphi) $ where the two
$ {\cal T} $-violating states can be adiabatically connected with each
other by changing $ \varphi $ from $ + \chi $ to $ - \chi $.
The spatial variation of $ \varphi
$ along the interface (with coordinate denoted as
$ x' $) is described by a generalized Ferrel-Prange equation which
is identical to a Sine-Gordon equation for standard Josephson
junctions ($ g(\varphi) = \sin \varphi $),

\begin{equation}
\frac{\partial^2 \varphi}{\partial x'^2} = \lambda_J^{-2}
g(\varphi),
\end{equation}

\noindent
with $ \lambda_J = (\Phi_0 c/ 8 \pi^2 d I_c)^{1/2}$ as the
characteristic
length scale ($ d $: the effective magnetic width of
the interface \cite{TINK}). As in conventional interfaces
the spatial variation of $ \varphi $
yields a finite local magnetic flux $ \phi(x') = (\Phi_0/ 2 \pi )
\partial \varphi(x') / \partial x' $. The integrated magnetic
flux between two points, $ x'_2 $ and $ x'_1 $, on the interface is
given by
$ \Phi = (\varphi(x'_2) - \varphi(x'_1)) \Phi_0 / 2 \pi $.

As a consequence, a spatial variation of the interface properties
can lead to the
occurrence of magnetic field distributions.
In particular, at the border between two different
homogeneous interface segments, (a) and (b),
characterized by their specific values of
$ \varphi_0 $ (denoted as $ \varphi_0^{(a)} $ and $ \varphi_0^{(b)} $,
respectively),
we find a continuous change of $ \varphi $ from
$ \varphi_0^{(a)} $ to $ \varphi_0^{(b)} $ on a length scale
$ \lambda_J $. On distances much larger than $ \lambda_J $
from the border of the segments $ \varphi $ is $ \varphi_0^{(j)} $
in segment $ (j) $.
The flux associated with the variation of $ \varphi $
is located in the vicinity of the border and is
$ \Phi = (\varphi_0^{(b)} - \varphi_0^{(a)})
\Phi_0 /2 \pi $. Obviously $ \Phi $ is not an integer (half-integer)
multiple of $ \Phi_0 $, but can have an arbitrary value.
It cannot be determined by simple topological arguments,
but depends on the specific properties of the interface.
Because the phase twist at the border between
segment (a) and (b) corresponds to a winding of the phase
$ \varphi_d $
by a fractional multiple of $ 2 \pi $,
this flux line may be considered as a {\it fractional vortex}
\cite{SRU}. Similarly fractional vortices can occur on a homogeneous
interface due to the twofold degeneracy of the $ {\cal T} $-violating
interface state. Two types of domains with $ \varphi_0 = + \chi $ and
$ - \chi $, respectively, are possible.
If present on the same interface, they are separated by domain
``walls''. There we find a kink of the width $ \lambda_J $ in
$ \varphi $ connecting
the two values of $ \varphi_0 $
and yielding a flux $ \Phi= \pm \chi \Phi_0 / \pi $
(see also \cite{SRU}).

We now propose an experiment to test our
picture. Let us consider a thin film consisting of two parts with
orientations of their crystalline axes different by $ 45^o $.
The boundary shall
be circularly curved (see Fig.3). The point P would correspond to
an interface with angle $ \theta = \pi/4 $ and $ \theta $
changes continuously as we turn away from P. Hence, along the
interface we scan through a certain range of angles $ \theta $.
If time-reversal symmetry were conserved everywhere, we would see a
vortex with $ \Phi = \Phi_0/2 $ at P, because P separates two
interface segments where one has $ \varphi_0 = 0 $ and the other
$ = \pi $ \cite{GESH,MILL}. The field distribution of the vortex
becomes more localized if we
cool the system because the penetration depth $ \lambda_J $
along the interface
shrinks with lowering temperature. On the other hand, if our
scenario is correct we expect that the phase $ \varphi_0 $ varies
continuously from 0 to $ \pi $ near P within a certain range of
$ \theta $.
This range extends with lowering temperature (see Fig.2),
such that the resulting field distribution becomes more extended
further as the system
is cooled, while the total flux always equals to $ \Phi_0/2 $.
The observation of the field
distribution by a (magnetic) microscope should allow to
prove or disprove our scenario based on these
qualitative properties.

In summary, we pointed out that important properties
of an interface between d-wave or other
unconventional superconductors result from
the intrinsic phase structure of the
pairing wave function $ \psi({\bf k}) $ which cannot
be found in conventional (s-wave) superconductors.
While some interface
coupling effects favor a $ {\cal T} $-invariant state,
in the bulk rather a
$ {\cal T} $-violating state is prefered.
In the competition between these
two tendencies it is important that
the interface coupling effects can be arbitrarily small
by choosing the interface geometry appropriately, i.e.,
the crystalline orientation of the two
connected superconductors. Thus,
proximity-induced order parameter components can combine with
the bulk order parameter to form a $ {\cal T} $-violating
interface state. We emphasize that such conditions
cannot be satisfied in conventional (s-wave)
superconductors, in general.
Therefore the proximity-induced time-reversal breaking interface
state is a consequence of unconventional superconductivity.
For the sake of simplicity we restricted ourselves
to the case of a system with tetragonal
crystal field symmetry. We note here, however, that
a slight orthorhombic distortion of
the lattice (as it is present in real materials) does not
change our conclusions qualitatively. It leads
to some minor modifications which are beyond the
scope of this letter and will be discussed
elsewhere in detail.

We are grateful to P.A. Lee, T.M. Rice, K. Ueda, A. Furusaki, Y.B.
Kim, D.K.K. Lee, C. Bruder
and D. Scalapino for helpful and stimulating discussions.
We acknowledge financial supports by Swiss Nationalfonds (M.S.)
and by Ministry of Education, Science and Culture of Japan (K.K.).
M.S. would like to thank the Institute for Solid State Physics of the
University of Tokyo and the University of Tsukuba
for their hospitality during the time when this work has been
finished.

\begin{figure}
\caption{Interface between d-wave superconductor A and B.
The shading on either side indicates the direction of
the crytalline $ x $-axis (the $ z $-axis points out of
the plane).}
\end{figure}

\begin{figure}
\caption{Schematice phase diagram in terms of temperature and crystal
orientation of side B. The shaded region denotes the $ {\cal
T} $-violating phase separating the 0- and the $ \pi $-junction phases.
Within the shaded region $ \chi $ changes continously.}
\end{figure}

\begin{figure}
\caption{Schematic arrangement of superconducting films
for a test-experiment. The direction of shading indicates
the crystalline orientation. The region A may be a d- or an s-wave
superconductor, while B must have d-wave symmetry. The point
P corresponds to the angle $ \theta = \pi/4 $ and $ \tilde{\theta}
= \pi/4 - \theta $.}
\end{figure}

\end{document}